# The Dynamics of Health Behavior Sentiments on a Large Online Social Network

Marcel Salathé* [1,2,3,] Duy Q. Vu [4], Shashank Khandelwal [1,2], David R. Hunter [4]

*1 Center for Infectious Disease Dynamics, Penn State University*

*2 Department of Biology, Penn State University*

*3 Department of Computer Sciences and Engineering, Penn State University*

*4 Department of Statistics, Penn State University*

**\*Corresponding author:**

Marcel Salathé, salathe@psu.edu

W-251 Millennium Science Complex, University Park, PA 18062, USA

Tel: (814) 867-4431

ABSTRACT

Modifiable health behaviors, a leading cause of illness and death in many countries, are often driven by individual beliefs and sentiments about health and disease. Individual behaviors affecting health outcomes are increasingly modulated by social networks, for example through the associations of like-minded individuals - homophily - or through peer influence effects. Using a statistical approach to measure the individual temporal effects of a large number of variables pertaining to social network statistics, we investigate the spread of a health sentiment towards a new vaccine on Twitter, a large online social network. We find that the effects of neighborhood size and exposure intensity are qualitatively very different depending on the type of sentiment. Generally, we find that larger numbers of opinionated neighbors inhibit the expression of sentiments. We also find that exposure to negative sentiment is contagious - by which we merely mean predictive of future negative sentiment expression - while exposure to positive sentiments is generally not. In fact, exposure to positive sentiments can even predict increased negative sentiment expression. Our results suggest that the effects of peer influence and social contagion on the dynamics of behavioral spread on social networks are strongly content-dependent.

Social networks play an important role in affecting the dynamics of health behaviors and the associated diseases (*1-3*), but identifying the main drivers of health behavior spread in social networks has been challenging. The observation that health behavior dynamics follow the patterns of social contacts - e.g. that behaviors are often clustered (*4*, *5*) and positively assorted at the dyadic level (*6*, *7*) - can be explained by multiple processes, the two most prominent being homophily and social influence. The homophily hypothesis posits that social contacts are a product of likemindedness, whereas the social influence hypothesis posits that likemindedness is a product of social contacts. Measuring and distinguishing between the effects of homophily and social influence can be difficult in observational studies (*6*, *8*, *9*), but is important for the development of health behavior intervention strategies. Vaccination behavior is a prime example of a health behavior shaping disease dynamics: outbreaks of disease preventable disease are more likely if overall vaccination rates decline (*10*), or if vaccination refusal is clustered in local communities (*11*, *12*). The continuously evolving public concern about vaccines despite the overwhelming scientific evidence on the safety of vaccines reflect the need for an increased understanding on how such sentiments spread over time (*13*).

Studying the dynamics of health behaviors on social networks can also be resource-intensive because social network data must often be inferred indirectly, and many health behaviors are complex and thus difficult to quantify. In recent years, online social media services have emerged as novel data sources where short messages are publicly shared, allowing for a detailed picture of the flow of information from person to person in large-scale networks. We have conducted a study to investigate the temporal dynamics of a readily quantifiable health sentiment—the intent to get vaccinated against a novel pandemic virus— on an online social network involving more than 100,000 people, and more than 4 million directed relationships among them. The health sentiment dynamics captured on this network are given by time-stamped messages published by the online social network users, retrospectively classified as expressing positive, neutral or negative sentiments about the intent to get immunized with pandemic influenza H1N1 vaccine (*7*). Although not directly measuring the health behavior, the data were shown to explain a large fraction of the spatial variance in CDC-estimated influenza A H1N1 vaccination rates. Insofar as the dynamics of these sentiment have shaped the dynamics of the health behavior, we are interested in the factors affecting the spread of health sentiments in the social network.

The data were collected from the online social networking service Twitter (www.twitter.com), where users post short messages (so-called 'tweets') of up to 140 characters that are then broadcast to their followers. Follower relationships are directional—if user A chooses to follow user B, user A will receive messages from user B, but user B will not receive messages from user A. In this case, we call user A a follower of user B, and user B a followee of user A (although followees are sometimes referred to as 'friends' in the media, we prefer the term followee because it more clearly conveys the direction of the relationship). Nevertheless, user B may also choose to follow user A, in which case a bidirectional relationship is established, and both users will receive messages from each other. An application programming interface (API) provided by Twitter allows for the collection of tweets matching a given set of requirements (e.g., containing a keyword), as well as the collection of follower and followee relationships among users. After data collection, machine learning algorithms were employed to label tweets as negative, positive or neutral with respect to the intent of getting vaccinated against influenza H1N1. We used an ensemble method combining a naive Bayes and a maximum entropy classifier with an accuracy of 84.29%. The full methodology is described in Salathé & Khandelwal 2011 (*7*).

In order to identify significant contributors to the likelihood that a user in the social network will express an opinionated (i.e., positive or negative) sentiment in the future, we use an approach that estimates the individual effects of numerous covariates related to the past sentiment expression behavior of users and social contacts as well as the structure of their social network neighborhood (Figure 1). We base our estimates on only the final 45 days of the data collection time period in order to ensure that they are based on a maximally accurate network representation: Network relationships could only be captured once a user had been identified as messaging about H1N1 vaccination, so cumulative network information improves toward the end of the data collection period. We associate two counting processes, $N_i^+(t)$ and $N_i^-(t)$, with each user *i* to count the number of positive and negative messages that the user has sent by time *t* (*14*). This results in multivariate counting processes $N^+(t) = [N_i^+(t), ..., N_i^+(t)]$ and $N^-(t) = [N_i^-(t), ..., N_i^-(t)]$, where *n* is the number of users in the network. By a mathematical result called the Doob-Meyer theorem (*14*), each of these (random) counting processes can be decomposed into an integrated conditional intensity process (the signal) and a random process called a martingale (the noise). We denote the conditional intensity functions for positive and negative tweeting events by user *i* as $\lambda^+(i, t \mid \beta^+, H_{t-})$ and $\lambda^-(i, t \mid \beta^-, H_{t-})$, where $H_{t-}$ is the network right before time *t*, and $\beta^+$ and $\beta^-$ are vectors of parameters.

Specifically, our models for the intensity functions $\lambda^+(i,t \mid \beta^+, H_{t-})$ and $\lambda^-(i,t \mid \beta^-, H_{t-})$ are Cox proportional hazards models (*15*), taking the form

$$\lambda^+(i,t \mid \beta^+, H_{t-}) = \lambda_0^+(t) \exp[\beta^+ \cdot s(i, H_{t-})] \qquad (1)$$

(similarly for $\lambda^-$). Here $s(i, H_{t-})$ is the covariate vector, which we discuss below, of node *i* based on the network history $H_{t-}$. In our model, each of the network covariates is multiplied by a corresponding element of one of the beta vectors, much like covariates in a regression model are multiplied by regression coefficients. Hence, the statistical significance of the estimated beta coefficients and their signs tell us how the corresponding covariates predict sentiment expression after correcting for all other covariate effects. We use exactly the same covariates in both models even though the coefficient vectors are different. The network covariates as summarized in Figure 1 capture a number of important aspects of network history $H_{t-}$ thought to be relevant for the dynamics of sentiment expression. A detailed description of all the covariates, along with a full list of the corresponding coefficient estimates and their p-values, is given in the Supplementary Material. Although we do not discuss them in the current paper, alternative methods for modeling $\lambda^+(i,t)$ exist. For instance, Vu et al. (*16*) discuss the so-called Aalen additive model for a similar situation, in which the effects of the covariates $s(i, H_{t-})$ are additive, rather than multiplicative, and the coefficients $\beta^+$ and $\beta^-$ may be assumed to change over time.

The coefficient vector $\beta^+$ in model given by equation (1), along with the vector $\beta^-$ corresponding to the analogous model for negative tweeting intensity, is estimated using maximum partial likelihood. This is standard practice for Cox proportional hazards models, though the computations are difficult in the present case due to the large size of the dataset. Thus, we employ computational innovations as outlined in Vu et al. (*17*). Using standard statistical theory, we may also obtain standard error estimates for each estimated coefficient, from which we may construct confidence intervals for each coefficient. However, because of the possible misclassification of the sentiment expressed in each tweet by the automatic classifier we employ, we do not base our statistical inferences on the single set of confidence intervals obtained from the dataset. Instead, we employ a series of random reclassifications of each tweet (the four categories being positive, negative, neutral, or unrelated to vaccination), based on a smaller set of test tweets used for calibration and using a method we detail in the Supplementary Material. In all, 200 different random reclassifications of every tweet are employed, and each such reclassification leads to a new realization of the network to which we apply our statistical estimation method. The resulting profile of 200 95% confidence intervals for every individual coefficient allows us to examine, in aggregate, the direction of each covariate's effect as well as its robustness against the misclassifications inherent in the automatic classification process. Examples of these sets of confidence intervals are presented in Figures 2 and 3 (with more given in the Supplementary Material).

Because our main interest is in assessing the effects of homophily and social contagion on the health sentiment dynamics in the network, we would like to measure the effects of both how many opinionated people a user is *connected* to, as well as how many opinionated messages a user is *exposed* to. These two effects are often confounded because on average, the more people a user is connected to, the more messages a user is exposed to. We therefore define covariates that separate these two effects as much as possible. A further important consideration is that users cannot simply be classified as positive or negative in their overall opinions because over the course of time they might have expressed different sentiments in numerous tweets. To address this issue, each followee is weighted by the fraction of opinionated (positive or negative) tweets he or she makes. The following paragraph gives precise definitions of these three positive-sentiment covariates as employed by the vector *s(i, $H_{t-}$)* of the model given by equation (1). The three corresponding negative-sentiment covariates are defined similarly. The full set of covariates, of which there are 24 in our full model, is explained in the Supplementary Material section.

In order to measure the extent to which a user is connected to people expressing positive or negative sentiments, we define the opinionated neighborhood size of a user to be the number of followees. The corresponding covariate, $f_1^+(i,t)$ as indicated in Figure 1, is defined as

$$f_1^+(i, t) = \sum_{j \in F(i,t)} \frac{N^+(j,t)}{N^a(j,t)}, \qquad (2)$$

where $F(i, t)$ is the set of followees of $i$ at time $t$ and $N^+(j, t)$ and $N^a(j, t)$ are, respectively, the number of positive tweets and the total number of vaccination-related tweets (positive, negative, or neutral, but excluding any tweets not related to H1N1 vaccination) made by $j$ before time $t$. We take the opinionated *reciprocal* neighborhood fraction of a user to be the proportion of followees that are reciprocal (i.e., who are also followers), weighted by the positivity fraction. The corresponding covariate, $f_5^+(i,t)$ in Figure 1, is defined as

$$f_5^+(i, t) = \frac{1}{f_1^+(i,t)} \sum_{j \in F(i,t)} \frac{N^+(j,t)}{N^a(j,t)} Y_{ji}(t), \qquad (3)$$

where $Y$ is the adjacency matrix of the network at time $t$ and thus $Y_{ji}(t)$ is the indicator that $j$ follows $i$ at time $t$. Finally, we define the average opinionated exposure intensity to be the weighted number of opinionated tweets by followees, normalized by the sum of the weights (to minimize the confounding with $f_1^+(i,t)$ as mentioned above). The corresponding covariate is

$$f_2^+(i,t) = \frac{1}{f_1^+(i,t)} \sum_{j \in F(i,t)} \frac{N^+(j,t)}{N^a(j,t)} N^+(j,t). \qquad (4)$$

We focus our attention on the six coefficients corresponding to the covariates described above, i.e., $f_1^+(i,t)$, $f_2^+(i,t)$, and $f_5^+(i,t)$ and their corresponding negative-sentiment covariates. We do not study the remaining 18 coefficients in the model with the same level of detail, both for the sake of simplicity and because our interest lies primarily in those effects that relate directly to social contagion. However, it is important that the other statistics, all of which are explained in the Supplementary Material, are included in the model, since this means that the six coefficients we discuss are estimated after accounting for the effects of all of the other statistics. For instance, we account for possible triangle-based clustering effects by including terms for average number of shared followers (of followees) and average number of shared followees (of followees); as we mention below, these terms control for some types of homophily.

The results are summarized in Figures 2 and 3. Generally, larger opinionated neighborhood sizes have an inhibitory effect on the expression of opinionated sentiments (Figure 2, A-D): While both larger positive and larger negative neighborhood sizes have the expected inhibitory effect on the expression of the opposite sentiments (i.e., negative and positive, respectively), they also predict diminished expression of that same sentiment. If we look at the opinionated reciprocal neighborhood size (Figure 3), we see that the effects are content-dependent, i.e., the effects are different for negative and positive sentiments. On one hand, larger positive reciprocal neighborhood sizes do not generally have a significant predictive effect on the rate of expressing opinionated sentiments. On the other hand, increasing negative reciprocal neighborhood size has the expected effect of increasing the likelihood of expressing a negative sentiment, and decreasing the likelihood of expressing a positive sentiment. Finally, the predictive effects of opinionated exposure intensity are also content-dependent (Figure 2, E-H). While a range of outcomes are observed in the 200 network realizations obtained via reclassifying each tweet's sentiment (as explained earlier), there is a sizable fraction of outcomes that show unexpected effects. In particular, in a substantial fraction of cases, being exposed to an increased intensity of positive tweets is predictive of increased intensity of negative sentiment (Figure 2G), as well as decreased intensity of positive sentiment (Figure 2H). Finally, the past expression of a sentiment by an individual predicts an increased propensity for that individual to express that same sentiment again, a finding that is very consistent across all 200 network realizations (see Figure S1).

It is worthwhile to consider these results in the context of what the statistics are expected to measure. Our main interest is in identifying the extent to which social contagion and homophily drive sentiment dynamics within the social network. In an observational study like the present study, causality cannot be established. Furthermore, disentangling effects of homophily and contagion is notoriously hard (*8*) because they are often confounded. Our approach tries to minimize these issues as much as possible. We use the term social contagion to mean the extent to which exposure to a given sentiment is predictive of future expression of that sentiment. Previous studies have focused on binary outcomes such as the adoption (vs. non-adoption) of a service (*6*, *18*), and have measured exposure as the number of social contacts that have adopted the service previously. Our methodology allows us to consider more complex measures of exposure: For instance, in the present analysis we measure both the number of social contacts expressing a given sentiment as well as the intensity with which the sentiment is expressed. Thus, both the opinionated neighborhood size as well as the average opinionated exposure intensity relate to social contagion as defined above. Homophily, on the other hand, is assessed by the opinionated *reciprocal* neighborhood size of a user, i.e., the weighted number of reciprocal followees, or followees who are also followers of that user.

The finding that the opinionated neighborhood size generally has an inhibitory effect on the likelihood of expressing any opinionated sentiment (Figure 2, A-D) is difficult to interpret in the context of a standard contagion framework, because contagion is normally associated with spread, rather than inhibition. For example, it makes intuitive sense that a larger number of negative followees should lead to a reduction in the expression of positive sentiments. The finding that it also leads to a reduction in the expression of negative sentiments is harder to interpret, but nevertheless agrees with the general pattern of inhibition. When looking at the average opinionated exposure intensity (Figure 2, E-H), a different picture emerges. The results are rather sensitive to misclassification of the messages, but the most stable result (64% of all network realizations, Figure 2E) is that increased average negative exposure intensity does predict increased negative sentiment expression, in line with the expectation of social contagion.

Surprisingly, the second most stable result (44.5% of all network realizations, Figure 2G) is that the average positive exposure intensity does also predict increased negative sentiment expression. Equally surprisingly, the third most stable result (33.5% of network realizations, Figure 2H) is that higher average positive exposure intensity predicts decreased positive sentiment expression. Taken together, the results suggest that exposure to negative sentiment is contagious—by which we merely mean predictive of future negative sentiment expression—while exposure to positive sentiments is generally not. They also suggest that exposure to increased intensity of opinionated sentiments has on balance led to increased negative sentiment expression and decreased positive sentiment expression, overall favoring the spread of negative vaccination sentiments.

The lack of detailed information about the users prohibits us from assessing manifest homophily, and our analysis is thus subject to the problem of latent homophily which is generally confounded with contagion (*8*). We assess homophily with the opinionated *reciprocal* neighborhood size of a user, which is the weighted number of reciprocal followees (i.e., followees who are also followers of that user). Bidirectional follower relationships mean that two users are interested in receiving messages from each other, an indicator that the users share the same interests, and thus of homophily. To further reduce the confounding effects of homophily and contagion, our model contains covariates for the number of shared followees and followers. These covariates are expected to control for latent homophily to a certain extent, since homophily is known to manifest itself in network clustering (*8*, *19*). Our findings suggest that the effects of homophily, insofar as we can measure it, are content-dependent: the positive reciprocal neighborhood size does generally not have significant effects (Figure 3, C and D), while increasing negative reciprocal neighborhood size has the expected effects of predicting decreased positive and increased negative sentiment expression (Figure 3, A and B). This finding further contributes to favoring the spread of negative vaccination sentiments.

Overall, the finding that the effects of various network covariates are strongly content-dependent suggests that a standard contagion framework might be too constrained to understand the health sentiment dynamics occurring on this network. By standard contagion framework, we mean the conceptual idea that increased exposure to any given agent (whether biological or social) will lead to an increased transmission - and predict an increased adoption - of that agent. In such a framework, the expectation is that there is a positive relationship between exposure and the consequent adoption of whatever it is individuals are exposed to. In our data, the only effect that corresponds to this pattern is that increased negative exposure intensity does predict increased negative sentiment expression. All the other results suggest that increased exposure predicts either a decrease of the same sentiment expression or an increase of the opposite sentiment expression.

From a public health perspective, the results raise some questions about the design of health behavior communication strategies. In particular, the notion that increased positive exposure intensity predicts increased negative sentiments could be of great concern if this turns out to be a consistent finding in future studies, since it would indicate that the level of positive messaging needs to be assessed carefully. Equally worrisome is the notion that the identified effects overall seem to favor the spread of negative sentiments, but not the spread of positive sentiments. This suggests that increased attention should be given to the prevention and control of negative sentiments (particularly if based on rumors, misinformation, misunderstandings etc.). The increased ability to measure the dynamics of sentiments on online networks generates opportunities to dramatically reduce the time lag between communication strategies and the assessment of the effects of those strategies.

The study framework has a number of limitations that need to be taken into account when assessing its applicability. First, our study design has been set up to catch expression of sentiments only (rather than actual vaccination behavior), but users might have been affected by exposure to sentiments from social contacts without ever expressing these sentiments themselves. For example, a user exposed to many negative messages may have been influenced and adopted a negative stance on H1N1 vaccination, but the user might not consequently have expressed that opinion in the network. Thus, a substantial fraction of actual contagion may have gone unnoticed. Conversely, peer pressure effects may have driven users to express a certain sentiment online even though they personally hold a different opinion (and behave differently from what one would expect based on the expressed sentiment), leading to false positives. Future research should address the question to what extent health sentiments expressed online overlap with actual health behaviors. Moreover, our study design ignores the possibility that follower relationships may have been established because users already share the same opinion on vaccination. While it is not unlikely that vaccination sentiments can be a contributor to establishing follower relationships, we believe that overall it had a small effect in the short period of time on which our analysis is based. Finally, the content of short messages like the ones studied here is subjective and open to interpretation by the reader of the message. Given the sometimes strong dependency of the effect on network realizations, this is an important problem that needs to be addressed in the future.

The dynamics of sentiments and behaviors on social networks is of great importance in many fields concerning human affairs (*20*), and particularly also in the health domain. There is an increased understanding that modifiable health behaviors are a key contributor to health outcomes (*21*), and that health behavior modification might be a key strategy to control major public health issues, both from the perspective of prevention (vaccination, smoking cessation, diet modification, etc.) and treatment (adherence to treatment plans, antibiotic overuse, etc.) strategies. The rapid worldwide adoption of online social network services means that an increasing fraction of (mis-)information diffusion is occurring on these networks. The methods and findings presented here are a small step towards an increased understanding of these dynamics, demonstrating both the promise and the challenges associated with these large and often unstructured data sets. In addition to online experiments (*22*, *23*), analysis of large-scale, high-resolution observational data will provide a much better picture of the dynamics of health behavior diffusion on social networks.

**ACKNOWLEDGMENTS**

This work is supported by a Society in Science - Branco Weiss Fellowship to Marcel Salathé, and by the Office of Naval Research (ONR grant N00014-08-1-1015) and the National Institutes of Health (NIH grant 1R01GM083603).

**FIGURE LEGENDS**

**Figure 1:** Illustration of covariates related to past sentiment expression behavior of users and social contacts as well as the structure of their social network neighborhood. Nodes represent users in the social network (the gray node represents the focal user), arrows represent the follower relationships, and numbers inside of nodes represent sentiment expression history (numbers of positive and negative sentiments expressed at given time; neutral sentiments are also counted, though we do not depict them here). The direction of the arrows represents the direction of information flow. Covariates $f_1$, $f_2$, and $f_5$ are explained in the article; the remaining covariates are explained in the Supplementary Material. For instance, the figure indicates that $f_1$ relates to the number of followees of the focal user and $f_2$ relates to the number of tweets these followees make, whereas $f_5$ counts reciprocated follower-followee relationships. Other covariates include information about the followers (as measured by $u_2$), the number of tweets made by the user ($u_1$), triangle-based covariates that measure certain types of clustering ($f_6$ and $f_7$), and numbers of follower and followees of the user's followees ($f_3$ and $f_4$). The figure illustrates that the values of these covariates may change with the advance of time (e.g. new tweets, new follower relationships, etc).

**Figure 2:** Estimated coefficients of covariates related to social contagion. Each panel shows the means (circle) and 95% confidence intervals (line) for 200 network realizations. The left column (A, C, E and G) are estimated coefficients for the likelihood of negative sentiment expression, the right column (B, D, F and H) are estimated coefficients for the likelihood of positive sentiment expression. The vertical dotted line is positioned at an estimated coefficient of zero (i.e. no effect). The percentage numbers in the top left corner of each panel indicate what fraction of the network realizations yielded statistically significant positive (green) or negative (red) coefficient estimates.

**Figure 3:** Like figure 2, but estimated coefficients of covariates related to homophily.

**FIGURE 1**

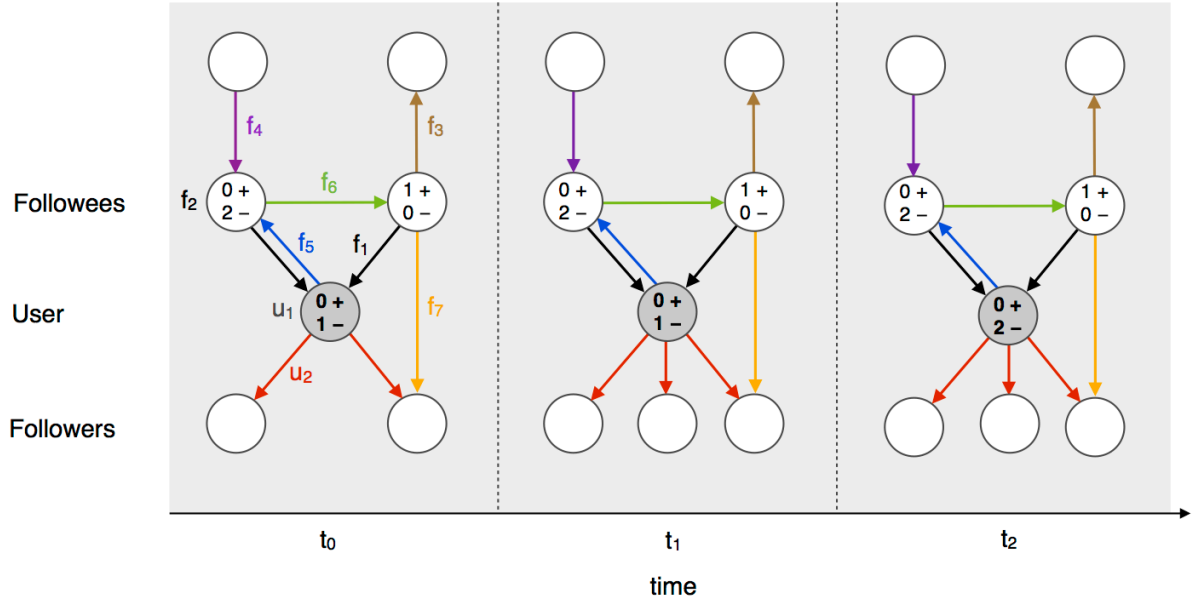

**FIGURE 2**

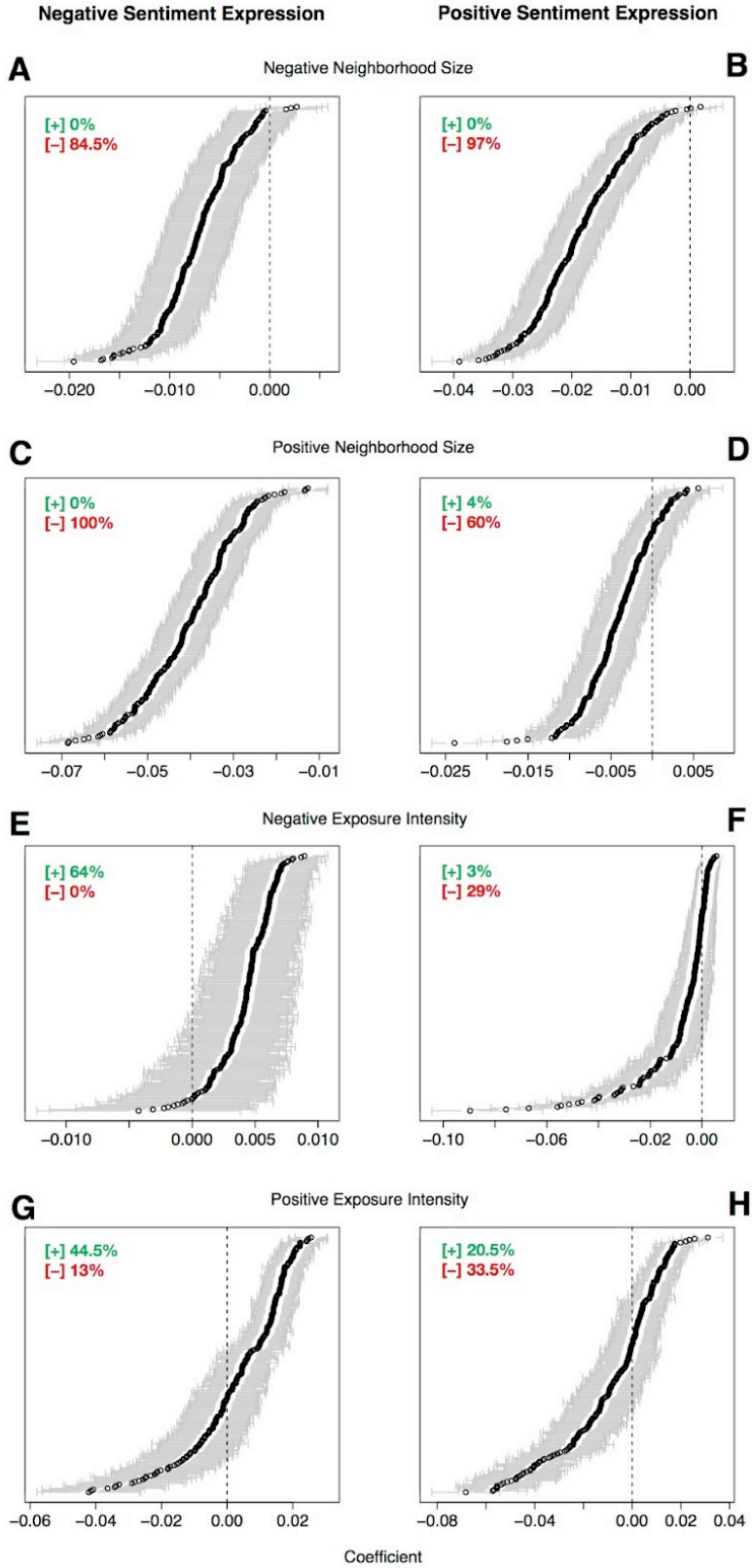

**FIGURE 3**

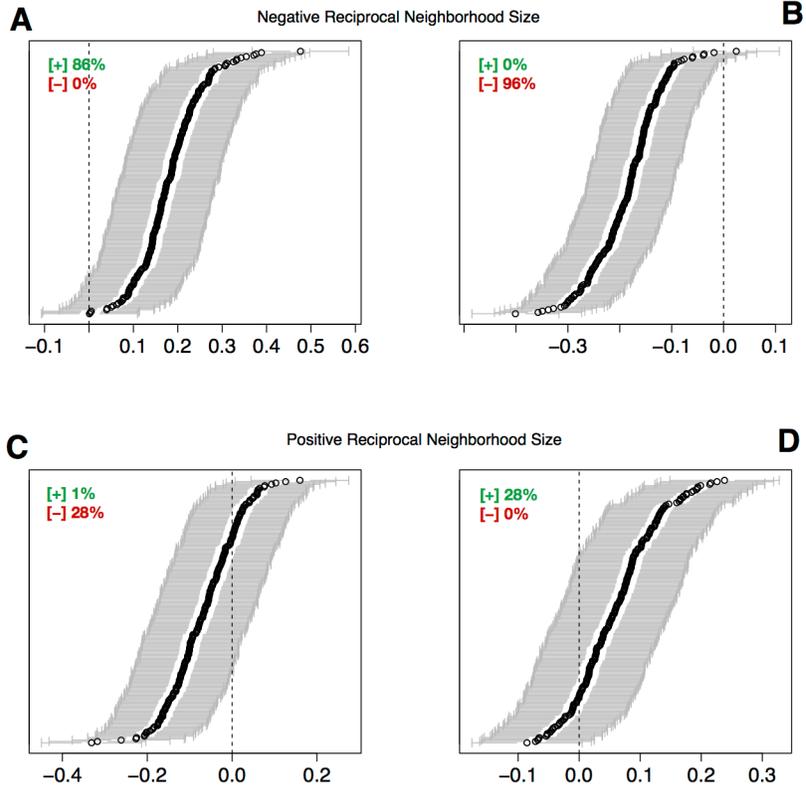

**SUPPLEMENTRAY MATERIAL**

Available at [www.salathegroup.com](www.salathegroup.com) (go to publications and find the reference to this article).